\documentclass[12pt]{article}
\pdfoutput=1
\usepackage{amssymb,amsmath,bm,bbm}
\usepackage{epsf}
\usepackage{epsfig}
\usepackage{afterpage}
\usepackage{longtable}
\usepackage{cite}
\usepackage{latexsym, mathrsfs}
\usepackage{color}

\newcommand{\be}{\begin{equation}}
\newcommand{\ee}{\end{equation}}

\newcommand{\beq}{\begin{equation}}
\newcommand{\eeq}{\end{equation}}
\newcommand{\bea}{\begin{eqnarray}}
\newcommand{\eea}{\end{eqnarray}}

\def\eps{\varepsilon}

\newcommand{\gsim}{\lower.7ex\hbox{$\;\stackrel{\textstyle>}{\sim}\;$}}
\newcommand{\lsim}{\lower.7ex\hbox{$\;\stackrel{\textstyle<}{\sim}\;$}}

\DeclareMathOperator{\re}{Re}
\DeclareMathOperator{\im}{Im}

\newcommand{\ord}{\mathcal{O}}

\newcommand{\bi}{\begin{itemize}}
\newcommand{\ei}{\end{itemize}}

\newcommand{\tev}{\, {\rm TeV}}
\newcommand{\gev}{\, {\rm GeV}}
\newcommand{\mev}{\, {\rm MeV}}

\newcommand{\eod}{\end{document}}

\def\cp{CP}

\newcommand{\DD}{D^0-\bar D^0}

\newcommand{\NP}{New Physics} 

\newcommand{\nn}{\nonumber}

\newcommand{\newsection}[1]{\section{#1}\setcounter{equation}{0}}

\begin{document}
\thispagestyle{empty}
\vspace*{-22mm}

\begin{flushright}
UND-HEP-08-BIG\hspace*{.08em}05\\
TUM-HEP-719/09\\
MPP-2009-36

\end{flushright}
\vspace*{10mm}

\begin{center}
{\Large {\bf\boldmath
\cp~Violation in $D^0 - \bar D^0$ Oscillations:\vspace{-1mm}\\
General Considerations and {Applications} \vspace{2mm}\\
to the Littlest Higgs Model with T-Parity
}}
\vspace*{15mm}

{\bf Ikaros I.\ Bigi$^a$, Monika Blanke$^{b,c}$, Andrzej J.\ Buras$^{b,d}$, Stefan Recksiegel$^b$} 
\vspace{7mm}

{\footnotesize
$^a$ {\sl Department of Physics, University of Notre Dame du Lac}\\
{\sl Notre Dame, IN 46556, USA}\vspace{1mm}

$^b$ {\sl Physik Department, Technische Universit\"at M\"unchen,
D-85748 Garching, Germany}\vspace{1mm}

$^c$  {\sl Max-Planck-Institut f{\"u}r Physik (Werner-Heisenberg-Institut), \\
D-80805 M{\"u}nchen, Germany}\vspace{1mm}

$^d$ {\sl TUM Institute for Advanced Study, Technische Universit\"at
   M\"unchen,
 \\ D-80333 M\"unchen, Germany}}
\vspace*{10mm}

{\bf Abstract}\vspace*{-1.5mm}\\
\end{center}

{\small
\noindent
The observed $D^0-\bar D^0$ oscillations provide 
a new stage in our search for New Physics in heavy flavour dynamics. The theoretical verdict on the observed values of $x_D$ and $y_D$ remains ambiguous: while they could be totally generated by Standard Model dynamics, they could also contain a sizable or even leading contribution from New Physics.  Those  oscillations are likely to 
enhance the observability of \cp~violation as clear manifestations of New Physics. 
We present general formulae for $D^0-\bar D^0$ oscillations,  concentrating on the case of negligible direct CP violation. In particular we derive a general formula for the time-dependent mixing-induced CP asymmetry in decays to a CP eigenstate and its correlation with the semileptonic CP asymmetry $a_\text{SL}(D^0)$ in $D^0(t)\to \ell\nu K$.
We apply our formalism to the Littlest Higgs model with T-parity, using the time-dependent CP asymmetry in {$D^0\to K_S\phi$} as an example. We find observable effects {at} a level well beyond anything possible with CKM dynamics. {Comparisons} with CP violation in the $K$ and $B$ systems offer an excellent test of this scenario {and reveal the specific pattern of flavour and CP violation in the $D^0-\bar D^0$ system predicted by this model}. 
We discuss a number of charm decays that could potentially offer an insight in the dynamics of CP violation in $D$ decays.
We also apply our formalism to $B_s-\bar B_s$ mixing.

\newpage
\setcounter{page}{1}

\hrule
\tableofcontents\vspace{5mm}

\hrule\vspace{5mm}

\newsection{Introduction}

To obtain a ``natural'' solution to the Standard Model's (SM) gauge hierarchy problem it has been  
conjectured that dynamics beyond it {have} to enter around the TeV scale. This problem has been further deepened by the fact 
that the SM has passed the test provided by electroweak parameters even on the
level of quantum corrections.  Little Higgs Models \cite{ArkaniHamed:2001ca,ArkaniHamed:2001nc} represent an intriguing response to this challenge. Rather than solving the gauge hierarchy problem they `delay the day of reckoning' to a higher scale. They provide scenarios where New Physics (NP) quanta can be produced at the LHC without creating conflicts with electroweak  constraints; at the same time they introduce many fewer additional parameters than SUSY or Extra Dimension scenarios. 

In order to avoid stringent electroweak precision constraints, one subclass of
those models introduces a discrete symmetry called T-parity \cite{Cheng:2003ju,Cheng:2004yc}, under which the new particles are odd and can therefore contribute only at the loop level. As a consequence a set of six T-odd `mirror' quarks needs to be introduced that are organised into three families \cite{Low:2004xc}. 
While constraints from flavour dynamics are {\em not} part of their 
motivation, this latter class of models is not of the Minimal Flavour Violation (MFV) \cite{Buras:2000dm,Buras:2003jf,D'Ambrosio:2002ex,Hall:1990ac,Chivukula:1987py} variety. They allow to construct 
connections between findings in high $p_t$ and flavour dynamics that have the
potential to be of practical use due to the relative paucity of their new
parameters:  ten in the quark flavour sector, among them three
  CP-violating phases.

If one had observed $D^0 - \bar D^0$ oscillations with 
$x_D > 1\% \gg y_D$, one would have had a strong prima facie case for NP enhancing $\Delta M_D$. 
Such a scenario has probably been ruled out now.  
The theoretical interpretation of the recent seminal discovery of $D^0 - \bar D^0$ oscillations 
with $x_D \sim y_D \sim (0.5 - 1) \%$ \cite{Aubert:2007wf,Staric:2007dt,Abe:2007rd,Barberio:2007cr} 
remains ambiguous \cite{Bigi:2000wn,Bianco:2003vb,Falk:2001hx,Falk:2004wg}: the observed size of $\Delta M_D$ and 
$\Delta \Gamma_D$ might {completely be due} to SM dynamics --- or $\Delta M_D$ could still contain sizable or even leading NP contributions. A breakthrough in our theoretical control over these 
quantities is required for resolving this issue on theoretical grounds. Barring that there are two possible interpretations of the present situation: (i) It is beyond our computational abilities to evaluate 
$\Delta M_D$ and $\Delta \Gamma_D$ with sufficient accuracy. (ii) It represents one example of nature being mischievous with us: $\Delta \Gamma_D$ is anomalously enhanced due to a violation of 
{\em local} quark-hadron duality; $\Delta M_D$ on the other hand is enhanced over the value expected in the SM due to the intervention of NP. There is no way that interpretation (ii) could be validated 
by theoretical arguments; yet we argue there is a straightforward course of action as outlined below.  

A priori Little Higgs models with T-parity could have generated considerably larger values of $\Delta M_D$ than observed;  yet in that case the accompanying weak phase in 
${\cal L}(\Delta C=2)$ had to be quite small due to constraints from $K_L\to \pi^+\pi^-$ decays 
\cite{Blanke:2007ee}. 
A forteriori they could generate the observed value or a significant fraction of it. The new feature now is that the $K_L$ constraints are diluted to a degree that large phases can emerge in 
${\cal L}(\Delta C=2)$. 
Their most striking experimental signature would be the observation of time dependent \cp~asymmetries  
already for Cabibbo allowed final states like $D^0 \to \phi K_S$ in {\em qualitative} --- albeit not quantitative --- analogy to $B_d \to \psi K_S$. 

The remainder of this {paper} is organised as follows. In Sect.\,\ref{sec:LH} we briefly recapitulate the basic features of Little Higgs models and describe the relevant ingredients of the Littlest Higgs model with T-parity (LHT) needed for our analysis. {Here we discuss in explicit terms the connection between $D$ and $K$ physics that is very transparent in the model in question.} Then in Sect.\,\ref{sec:DD}, we review the theoretical framework of $D^0-\bar D^0$ oscillations and discuss the possible LHT contributions, which we compare with the experimental evidence.  Sect.\,\ref{sec:CPV} is dedicated to a model-independent discussion of the effect of indirect CP violation on {$D$ decays}, where we derive in particular a correlation between the semileptonic asymmetry $a_\text{SL}$ and the time-dependent asymmetry in {$D^0\to K_S\phi$}. In Sect.\,\ref{sec:CPV-LHT} we apply the formalism of Sect.\,\ref{sec:CPV} to the LHT model and show the possible effects of LHT dynamics in CP violation in $D^0-\bar D^0$ oscillations. In Sect.\,\ref{sec:KB} we  
consider simultaneously the impact of LHT dynamics on CP violation in $D^0-\bar D^0$ oscillations, rare CP violating $K_L$ decays and CP violation in $B_s-\bar B_s$ mixing, measured by the CP asymmetry $S_{\psi\phi}$.
A brief summary and outlook is given in Sect.\,\ref{sec:sum}. 
{Finally using the formalism of Sect.\,\ref{sec:CPV} Appendix \ref{app:Bs} rederives the correlation between the CP asymmetry $S_{\psi\phi}$ and the semileptonic asymmetry {$a^s_\text{SL}$} relevant for $B_s-\bar B_s$ mixing. Appendix \ref{app:input} collects the input parameters used in our analysis.}

\newsection{\boldmath Little Higgs Basics}
\label{sec:LH}

\subsection{Generalities}

The Little Higgs class \cite{ArkaniHamed:2001ca,ArkaniHamed:2001nc}
comprises a large variety of \NP~models in which the Higgs boson appears as a pseudo-Goldstone boson of a spontaneously broken global symmetry. Gauge and Yukawa couplings break the global symmetry explicitly; however, Little Higgs models are constructed such that every single coupling preserves enough of the global symmetry to keep the Higgs boson massless. Only when more than one coupling is non-vanishing, the symmetry is broken completely and radiative corrections to the Higgs potential arise, being however at most logarithmically divergent at the one-loop level.

In order to achieve the cancellation of quadratically divergent contributions
from the top quark, the electroweak gauge bosons and the Higgs itself, a
common feature of all Little Higgs models is a set of new heavy weak gauge
bosons, scalars and a top partner $T$ at the TeV scale. In spite of the large
variety of existing models on the market, this common feature leads in many
cases to similar phenomenological implications. Most phenomenological analyses
therefore restrict their attention to the Littlest Higgs (LH) model 
\cite{ArkaniHamed:2002qy}.
This latter model, based on an $SU(5)\to SO(5)$ global symmetry breaking
pattern at a scale $f\sim 1\tev$, introduces in addition to the SM gauge and
matter fields the heavy gauge bosons $W_H^\pm$, $Z_H$ and $A_H$, the heavy top
partner $T$ and a scalar triplet $\Phi$. In the remainder of this paper, we
will restrict ourselves to this economical realization of the Little Higgs
concept. Reviews can be found in \cite{Schmaltz:2005ky,Perelstein:2005ka}.

\subsection{The Littlest Higgs Model with T-Parity}

When studying electroweak precision observables, it turns out that an
additional discrete symmetry, called T-parity \cite{Cheng:2003ju,Cheng:2004yc}, is needed in order to allow for
the new particles below the $1\tev$ scale. Under this symmetry, the SM
particles and the heavy top partner $T_+$ are even, while ${W_H^\pm}, Z_H, A_H$ and
$\Phi$ are odd. A consistent implementation of T-parity requires also the
introduction of mirror fermions --- one for each quark and lepton species --
that are odd under T-parity \cite{Low:2004xc}. In this manner the Littlest Higgs model with
  T-parity (LHT) is {born.} 

While the Littlest Higgs model without T-parity belonged to the MFV class of
models, implying generally small effects in flavour violating observables
\cite{Buras:2004kq,Buras:2006wk}, the mirror fermions in the LHT model
introduce new sources of flavour and 
\cp~violation \cite{Hubisz:2005bd,Blanke:2006xr}. Potentially large deviations from the SM and MFV
predictions in flavour changing neutral current processes can thus appear
\cite{Blanke:2006sb,Blanke:2006eb,Blanke:2007db,Blanke:2007ee,Blanke:2007wr,Blanke:2008ac,LHT-update}.
A brief review of these analyses can be found in \cite{Blanke:2007ww}.

Flavour mixing in the mirror sector is conveniently described by two unitary
mixing matrices $V_{Hu}$ and $V_{Hd}$, parameterising the mirror quark
couplings to the SM up- and down-type quarks 
\cite{Hubisz:2005bd,Blanke:2006xr}, respectively. With the $V_{Hd}$ matrix being parameterised in terms of three mixing angles and three complex phases as suggested in \cite{Blanke:2006xr}, the $V_{Hu}$ matrix, relevant for $\DD$ oscillations, is given by
\beq\label{basicr}
V_{Hu} = V_{Hd} V_\text{CKM}^\dagger\,.
\eeq
As the CKM mixing angles are experimentally found to be small and therefore
$V_\text{CKM}\simeq\mathbbm{1}$, we have $V_{Hu} \simeq V_{Hd}$. This relation will turn out to be important in what follows in order to understand the close connection of \cp~violation in the $K$ and $D$ meson systems. This issue is discussed in more detail in the next section (see also \cite{Blanke:2007ee,Blum:2009sk}).

For an extensive description of the LHT model we refer the reader to
\cite{Blanke:2006eb}, where also a complete set of Feynman rules has been
derived. We note that an error in the coupling of the $Z$ boson to the mirror
fermions has been pointed out in \cite{Goto:2008fj}, see also \cite{delAguila:2008zu,LHT-update}. This has however no impact on the present analysis, as the coupling in question does not appear in the one-loop diagrams contributing to $\Delta F=2$ processes.

\subsection[Connection between $D$ and $K$ Physics]{\boldmath Connection between $D$ and $K$ Physics}\label{sec:xi}

In \cite{Blum:2009sk} the connection between $D^0-\bar D^0$ and $K^0-\bar K^0$ mixing has been discussed within the framework of approximately $SU(2)_L$-invariant \NP. Due to the {connection between up- and down-type quarks in the SM through the CKM matrix}, in this scenario the \NP~contributions to $D^0-\bar D^0$ and $K^0-\bar K^0$ mixing are not independent of each other. This observation has been used in \cite{Blum:2009sk} to derive lower bounds on the \NP~scale in various \NP~scenarios, emerging if the experimental constraints on $D^0-\bar D^0$ and $K^0-\bar K^0$ mixing are applied to only the $(V-A)\otimes (V-A)$ contribution. 
One should keep in mind however that in models in which new operators contribute to $\Delta F=2$ processes the power of this approach is limited, as the various contributions interplay with each other and dilute the correlation in question. 

On the other hand the situation is promising in \NP~models with only SM operators, such as the LHT model. {In fact this model provides possibly the best example of the physics discussed in \cite{Blum:2009sk}.} While the following discussion {has been triggered by the} analysis of $\Delta C=2$ processes in the LHT model and uses the notations and conventions of \cite{Hubisz:2005bd,Blanke:2006eb,Blanke:2007ee}, it applies as well to all other \NP~scenarios with only SM operators. Similar to the LHT model, the flavour mixing matrices $V_{Hu}$ and $V_{Hd}$ parameterise the misalignment between the \NP~and the SM up- and down-type quarks, respectively, that are related via \eqref{basicr}. $D^0-\bar D^0$ oscillations are then governed by the combinations ($i=1,2,3$)
\beq
\xi_i^{(D)} = {V_{Hu}^{iu}}^* V_{Hu}^{ic}\,,
\eeq
while for $K$, $B_d$ and $B_s$ physics
\beq
\xi_i^{(K)} = {V_{Hd}^{is}}^* V_{Hd}^{id}\,,\qquad
\xi_i^{(d)} = {V_{Hd}^{ib}}^* V_{Hd}^{id}\,,\qquad
\xi_i^{(s)} = {V_{Hd}^{ib}}^* V_{Hd}^{is}\,,
\eeq
respectively, are relevant. By making use of \eqref{basicr}, we can now express $\xi_i^{(D)}$ through combinations of $V_{Hd}$ and CKM elements. Using the Wolfenstein parameterisation for $V_\text{CKM}$ and expanding in powers of $\lambda$, we find
\bea
\xi_i^{(D)} = {\xi_i^{(K)}}^*  &+& \lambda \left( |V_{Hd}^{is}|^2 - |V_{Hd}^{id}|^2 \right) 
 + \lambda^2 \left( A {\xi_i^{(d)}}^* - 2 \re(\xi_i^{(K)}) \right) \nn\\
&+& \lambda^3 \left( \frac{1}{2} (|V_{Hd}^{id}|^2-|V_{Hd}^{is}|^2) 
                           + A {\xi_i^{(s)}}^* 
                           + A \xi_i^{(s)} (\rho-i\eta) \right) + \ord(\lambda^4)\,.\label{eq:xi}
\eea

The following comments are in order:
\bi
\item
At leading order $\xi_i^{(D)} = {\xi_i^{(K)}}^*$, i.\,e. $D$ and $K$ physics are governed by the \emph{same} \NP~flavour structure. We note that the complex conjugation arises, as $|D^0\rangle =| \bar u c \rangle$ while $|K^0\rangle =| \bar sd \rangle$, and it will give rise {to} a sign difference in \cp~violating effects in the $D$ and $K$ systems.
\item
The correction to linear order in $\lambda$ is real, irrespective of the precise structure of $V_{Hd}$. However, as $\Delta C=2$ \cp~violation is governed by
\beq  
\im (\xi_i^{(D)})^2 = 2 \re  \xi_i^{(D)} \im \xi_i^{(D)}\,,
\eeq
corrections to the one-to-one correspondence between $D^0-\bar D^0$ and
$K^0-\bar K^0$ mixings will {appear}  already at $\ord(\lambda)$ in both \cp~conserving and violating observables. On the other hand the {$\Delta C=1$ effective Hamiltonian is}  governed by a single power of $\xi_i^{(D)}$, so that {direct \cp~violation} in rare $D$ and $K$ decays will be much more strongly correlated and deviations from the one-to-one correspondence will arise only at $\ord(\lambda^2)$.
\item
The order $\ord(\lambda^2)$ correction can be complex, provided that $\im\xi_i^{(d)}\ne 0$, i.\,e.\ that there are new \cp~violating effects in the $B_d$ system.
\item
At $\ord(\lambda^3)$ a complex correction arises due to the \cp~violation in
the CKM matrix, given by $i\eta$. This correction is non-vanishing also in the
limit of a real, i.\,e. \cp~conserving $V_{Hd}$, and {vanishes} 
only if $\xi_i^{(s)}=0$.
\ei

\newsection{\boldmath $D^0 - \bar D^0$ Oscillations\label{sec:DD}}

\subsection{Theoretical Framework}

The time evolution of neutral $D$ mesons is generally described by the Schr\"odinger equation
\beq\label{eq:DGL}
i\frac{\partial}{\partial t} \begin{pmatrix}D^0 \\ \bar D^0 \end{pmatrix} =
\addtolength{\arraycolsep}{3pt}
\left(\begin{array}{cc} M_{11}^D -\frac{i}{2}\Gamma_{11}^D &
 M_{12}^D -\frac{i}{2}\Gamma_{12}^D \\
 {M_{12}^D}^* -\frac{i}{2}{\Gamma_{12}^D}^* &
 M_{11}^D -\frac{i}{2}\Gamma_{11}^D 
\end{array}\right)\addtolength{\arraycolsep}{-3pt}
\begin{pmatrix}D^0 \\ \bar D^0 \end{pmatrix}\,.
\eeq
In the presence of flavour violation
\beq
 M_{12}^D \ne 0 \,, \qquad \Gamma_{12}^D \ne 0\,,
\eeq
and the mass eigenstates can be written as
\bea
| D_1 \rangle &=& \frac{1}{\sqrt{|p|^2+|q|^2}} \left( p | D^0 \rangle + q | \bar D^0 \rangle \right)\,,\\
| D_2 \rangle &=& \frac{1}{\sqrt{|p|^2+|q|^2}} \left( p | D^0 \rangle - q | \bar D^0 \rangle \right) \,,
\eea
where
\beq 
\frac{q}{p} \equiv \sqrt\frac{{{M_{12}^D}^*} - \frac{i}{2}{{\Gamma_{12}^D}^*}}{M_{12}^D - \frac{i}{2}\Gamma_{12}^D} \, ,
\label{qoverp}
\eeq
and we choose {the CP phase convention} 
\beq\label{eq:CPC}
\cp |D^0\rangle = + |\bar D^0\rangle\,.
\eeq

$D^0 - \bar D^0$ oscillations can then be characterised by 
the  normalised mass and width differences 
\beq
x_D \equiv \frac{\Delta M_D}{\overline \Gamma} \, , 
\qquad y_D \equiv \frac{\Delta \Gamma_D}{2\overline \Gamma} \, , \qquad 
\overline \Gamma = \frac{1}{2} (\Gamma_1 + \Gamma_2) \, ,  \label{eq:xDyD}
\eeq
with 
\bea 
\Delta M_D &=& M_1 - M_2 = 2 \re\left[ \frac{q}{p}\left( M_{12}^D - \frac{i}{2}\Gamma_{12}^D\right) \right] \nn\\
& = &  2 \re\sqrt{|M_{12}^D|^2 - \frac{1}{4} |\Gamma_{12}^D|^2 - i\re(\Gamma_{12}^D{M_{12}^D}^*)}\,,\label{eq:DM} \\
\Delta \Gamma_D &=& \Gamma_1 - \Gamma_2 = -4 \im\left[ \frac{q}{p}\left( M_{12}^D - \frac{i}{2}\Gamma_{12}^D\right)\right] \nn\\
&=& - 4 \im\sqrt{|M_{12}^D|^2 - \frac{1}{4} |\Gamma_{12}^D|^2 - i \re(\Gamma_{12}^D{M_{12}^D}^*)} \label{eq:DG}\,.
\eea

{The} attentive reader will note that our definitions 
in eqs.\ (\ref{eq:DM}), \eqref{eq:DG} follow the PDG conventions \cite{Amsler:2008zzb}, also adopted by the HFAG collaboration \cite{Barberio:2007cr}; for neutral kaons they 
lead to $\Delta M_K \cdot \Delta \Gamma_K < 0$.  
Note that if $|\Gamma_{12}^D| \ll |M_{12}^D|$, as appropriate for $B^0$ mesons, one would  recover the familiar 
expressions $\Delta M \simeq 2 |M_{12}^D|$, $\Delta \Gamma \ll \Delta M$. 

While $\Delta M_D$ and $\Delta \Gamma_D$ tell us nothing about \cp~symmetry, the ratio $q/p$ and the relative phase between $M_{12}^D$ and $\Gamma_{12}^D$,
{\be\label{eq:phi12}
\varphi_{12} =\frac{1}{2}\arg\left(\frac{M_{12}^D}{\Gamma_{12}^D}\right)\,,
\ee
{express} the \cp~impurity in the two mass eigenstates through $|q/p| \neq 1$ and/or $2\varphi_{12}\ne \{0,\pm\pi\}$.  We note that while the phases of $M^D_{12}$ and $\Gamma^D_{12}$ depend on the phase conventions chosen, $\varphi_{12}$ is phase convention independent and consequently an observable.}

The world averages based on data from BaBar, Belle and CDF read \cite{Barberio:2007cr}
\begin{gather}
x_D = 0.0100^{+0.0024}_{-0.0026} \, , \qquad y_D = 0.0076^{+0.0017}_{-0.0018} \, , \qquad 
\frac{x_D^2 + y_D^2}{2} \leq (1.3 \pm 2.7)\cdot 10^{-4} \label{eq:xyexp}\\
\left| \frac{q}{p}\right| = 0.86 ^{+0.17}_{-0.15} \label{eq:qpexp}
\end{gather}
In the limit of (approximate) \cp~symmetry $x_D$, $y_D > 0$ implies the \cp~{\em even} state to be 
slightly heavier and shorter lived than the \cp~{\em odd} one (unlike for neutral kaons).

While there is close to universal consensus that $D^0 - \bar D^0$ oscillations have been observed --- 
$(x_D,y_D) \neq (0,0)$ --- considerable uncertainty exists concerning the relative and absolute sizes 
of $x_D$ and $y_D$. In what follows we will use the experimental $1\sigma$
ranges for $x_D$ and $y_D$ from (\ref{eq:xyexp}).

No sign of \cp~violation has been observed yet: the value of $|q/p|$ is fully consistent with unity, and a 
time integrated \cp~asymmetry in $D^0 \to K^+K^-$ or $\pi^+\pi^-$ is bounded by about 
$(0.5 - 1) \%$. Yet the following should be kept in mind: (i) The experimental uncertainty on 
$|q/p|$ is still quite large. (ii) $D^0 - \bar D^0$ oscillation can induce a time integrated \cp~asymmetry 
$\simeq x_D\cdot {\sin}2\varphi_f$ or $y_D\cdot {\sin}2\varphi_{f}$ as described in more 
detail in Sect.\ref{sec:CPV}; with $x_D$ and 
$y_D$ bounded by about 0.01, such an asymmetry can hardly exceed 1\%; i.e. we have just entered 
a regime where one can realistically hope for an effect to emerge. 

A few technical remarks are in order to set the basics for our subsequent discussions. The 
phases of neither $M_{12}$ nor $\Gamma_{12}$ are observable {\em per se}, since they depend on the 
phase convention adopted for $\bar D^0$; their relative phase {$\varphi_{12}$} however is independent of that 
convention and represents an observable. The CKM matrix provides a very convenient phase convention for $M_{12}^D$, which we will adopt. In the SM $M_{12}^D$ as well as 
$\Gamma_{12}^D$ are real to a very good approximation; however this still leaves 
their signs to be decided. {While the} authors of \cite{Falk:2004wg} argue that in the SM
$(\Gamma_{12}^D)_{\rm SM}$ {likely} carries a relative minus sign with respect to $(M_{12}^D)_{\rm SM}$, {the data} on $x_D$ and $y_D$, assuming no NP contribution, imply
\beq
(M_{12}^D)_{\rm SM} \sim 0.012 \text{\,ps}^{-1}\,,\qquad (\Gamma_{12}^D)_{\rm SM} \sim 0.018 \text{\,ps}^{-1} \,,
\eeq
 i.\,e.\ a relative plus sign between dispersive and absorptive part of the off-diagonal mixing element. {In what follows we will therefore not make any assumption on the signs of $(M_{12}^D)_\text{SM}$ and $(\Gamma_{12}^D)_\text{SM}$.}

\subsection{LHT Contributions}

The leading LHT contribution to ${\cal L}(\Delta C =2)$ is given by the
standard $(V-A)\otimes(V-A)$ operator with its Wilson coefficient modified by
the exchanges of the mirror quarks and heavy gauge bosons
$W_H^\pm$, $Z_H$ and $A_H$ in the relevant box diagrams. While the heavy $T_+$ quark cannot contribute directly to the box diagrams in question, its mixing with the standard up-type quarks can generate tree-level flavour changing $Z$ couplings in the up quark sector, leading to a non-vanishing contribution to $D^0-\bar D^0$ mixing {\cite{Lee:2004me,Fajfer:2007vk}}. However, if the heavy $T_+$ state is quasi-aligned with the SM top quark, as required in order not to spoil the Little Higgs mechanism of collective symmetry breaking, these tree level contributions are found to be smaller by several orders of magnitude than the SM short distance contributions and therefore fully negligible.

Explicit expressions {for the T-odd contributions to $D^0-\bar D^0$ mixing}
can be found in \cite{Blanke:2007ee} and will not be repeated here. We only
recall certain properties of these formulae that are relevant for our
work:
\begin{itemize}
\item
 The LHT contribution depends on seven new real parameters and three complex 
phases that are constrained to some extent by the data on FCNC processes in
$K$ and $B$ systems, in particular by the observed CP violation in $K_L\to\pi\pi$
decays through the {relations (\ref{basicr}) and \eqref{eq:xi}} and by electroweak precision tests.
The hadronic uncertainties in this contribution originate in the matrix element
of the relevant $\Delta C=2$ operator between $D^0$ and $\bar D^0$ states.
This matrix element is parameterised by the $D$ meson decay constant $F_D$
and the parameter $\hat B_D$. It should be emphasised that these two 
parameters are known from lattice calculations with much higher precision
than the analogous quantities in the $B_d$ and $B_s$ systems. Therefore in
view of other uncertainties in our analysis, primarily related to long distance
contributions discussed next, it is justified to set $\hat B_D$ and $F_D$ to
their central values.
\item
The remaining two contributions are the SM box contribution and the genuine
long distance contribution connected with low energy QCD dynamics.
Whatever the nature and strengths of the SM contributions, they have 
nothing to do with the physics of  the  LHT model and are always
 present;  we will denote the sum of these two contributions to the off-diagonal
element of the $D^0-\bar D^0$ mixing matrix simply by $(M^D_{12})_{\rm SM}$.
 It  is real to an excellent approximation   \cite{Bigi:2000wn,Bianco:2003vb}.
\end{itemize}

{In summary, at} present we have two experimental constraints of rather moderate rigour --- 
$\Delta M_D$ and $\Delta \Gamma_D$ --- and some order of magnitude estimates for the 
SM contributions to $M^D_{12}$ and $\Gamma^D_{12}$; furthermore we can count on the complex phases 
of the latter to be so small that they can be ignored at present. Lastly we have only  
constraints on the LHT parameters. For all these reasons we can provide only more or less 
typical scenarios, which we construct in the following way.  

We find sets of LHT parameters consistent with the data outside charm dynamics and then 
compute $M_{12}^D$ from them. To this end we fix the \NP~scale to $f=1\tev$, implying masses of the heavy gauge bosons
\beq
M_{W_H, Z_H}= gf\sim 650\gev\,,\qquad  M_{A_H}=\frac{g'f}{\sqrt{5}}\sim 160\gev\,.
\eeq
While the heavy gauge boson masses are fixed by the choice of $f$, the mirror quark masses $m_H^i$ depend on additional free Yukawa coupling parameters $\kappa_i$. Therefore we vary the mirror quark masses over the range
\beq
300\gev \le m_H^i \le 1000\gev\,.
\eeq
Note that the masses of up and down mirror quarks in the same doublet are  approximately equal.
The parameter $x_L$, describing the mixing between the top quark and the heavy $T_+$, is fixed to $x_L=0.5$ in our analysis. While it does not enter $D^0-\bar D^0$ mixing directly, it is relevant for the constraints from $K^0-\bar K^0$ and $B_{d,s}^0-\bar B_{d,s}^0$ mixing.

{Setting
\bea
 M_{12}^D &=& (M^D_{12})_{\rm SM}+(M^D_{12})_{\rm LHT} \,,\label{eq:M12}\\
\Gamma_{12}^D &=& (\Gamma_{12}^D)_\text{SM} \,, 
\eea
we then ask what {\em real} values are required for 
$(M_{12}^D)_\text{SM}$ and $\Gamma_{12}^D = (\Gamma_{12}^D)_\text{SM}$ to reproduce a size for 
$\Delta M_D$ and $\Delta \Gamma_D$ that is compatible with the data. }

\begin{figure}
\center{\includegraphics[width=.7\textwidth]{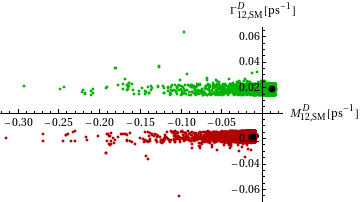}}
\caption{\it $(\Gamma^D_{12})_\text{SM}$ as a function of $(M^D_{12})_\text{SM}$.
The red (darker grey) and green (lighter grey) points correspond to the two
solutions when solving \eqref{eq:DM}, \eqref{eq:DG} for the poorly known SM
contribution. In this and in all subsequent plots, the thick black points
correspond to the SM case, i.e.\ the LHT contribution has been set to zero {and the SM contribution alone reproduces the experimental (central) values of $x_D$ and $y_D$.}
\label{fig:MD12SM-GD12SM}}
\end{figure}

The result of this procedure is shown in Fig.\ \ref{fig:MD12SM-GD12SM}, where we show $(\Gamma^D_{12})_\text{SM}$ as a function of $(M^D_{12})_\text{SM}$. As there are generally two solutions for the SM contribution, we determine both and show them as red (darker grey) and green (lighter grey) points, respectively, in this and all further figures. We observe that while for essentially all LHT parameter points $(\Gamma_{12}^D)_\text{SM}$ is consistent with theoretical estimates, for some points a very large negative $(M_{12}^D)_\text{SM}$ is needed. We have verified explicitly that those points do not coincide with the most spectacular effects discussed below and in the $K$ and $B$ physics observables discussed in \cite{Blanke:2006sb,Blanke:2006eb,Blanke:2007db,LHT-update}. As an example we show in Fig.\ \ref{fig:MD12SM-qp} $|q/p|$ as a function of $(M^D_{12})_\text{SM}$, with the deviation of $|q/p|$ from {unity} measuring the size of \cp~violating effects in $D^0-\bar D^0$ oscillations.

\begin{figure}
\center{\includegraphics[width=.7\textwidth]{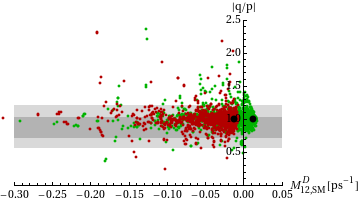}}
\caption{\it $|q/p|$ as a function of $(M^D_{12})_\text{SM}$. The red (darker grey) and green (lighter grey) points correspond to the two solutions for $((M_{12}^D)_\text{SM},(\Gamma_{12}^D)_\text{SM})$.\label{fig:MD12SM-qp}}
\end{figure}

\begin{figure}
\center{\includegraphics[width=.7\textwidth]{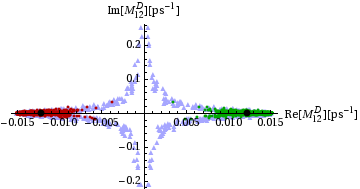}}
\caption{\it $\im(M_{12}^D)$ as a function of $\re(M_{12}^D)$. The red (darker
  grey) and green (slightly lighter grey) points fulfil all existing $K$ and
  $B$ physics constraints and correspond to the two solutions for
  $((M_{12}^D)_\text{SM},(\Gamma_{12}^D)_\text{SM})$, while for the light blue
  (grey) triangular points the constraint from $\eps_K$ has been omitted.
 \label{fig:ReMD12-ImMD12}}
\end{figure}
{In Fig.\ \ref{fig:ReMD12-ImMD12} we show the real and imaginary part 
of $M_{12}^D$, as defined in \eqref{eq:M12}. 
Again the red (darker grey) and green (slightly lighter grey) points fulfil all
existing $K$ and $B$ physics constraints and correspond to the two solutions
for $((M_{12}^D)_\text{SM},(\Gamma_{12}^D)_\text{SM})$, while for the light
blue (grey) triangular points the constraint from $\eps_K$ has been omitted. We
observe that even in the latter case, a strong correlation {between 
$\re(M_{12}^D)$} and $\im(M_{12}^D)$ appears. This is due to the
experimental constraints on $x_D$ and $y_D$ which enter by solving
(\ref{eq:xDyD}) for $((M_{12}^D)_\text{SM},(\Gamma_{12}^D)_\text{SM})$. {We note that very large values of $\im M_{12}^D$ {(Note the vastly different scales on the two axes!)} generally have to be compensated by an unnaturally large $(\Gamma_{12}^D)_\text{SM}$ in order to agree with the data.}
The additional constraint from}
$\eps_K\propto \im M_{12}^K$ results in the allowed red (darker grey) and 
green (slightly lighter grey) areas in the figure. We observe that points with {very}
large $\im M_{12}^D$ are now excluded, due to the correlation between $K$
and $D$ physics {discussed analytically in Sect.\ \ref{sec:xi}}. On the other hand we observe that almost the entire range of
CP-violating phases is allowed, although phases close to $\pm90^\circ$, or
equivalently {$\varphi_{12}=\pm45^\circ$}, appear to be unlikely. It should be
emphasised that independently of what fraction of the observed $\Delta M_D$ is
attributed to the SM contribution, a non-negligible phase
$\varphi_{12}$ can only come from NP, in our case from LHT contributions. This
makes it very clear that an observation of large mixing induced CP asymmetries
in $D$ decays which are governed by the phase {$\varphi_{12}$}, would be a clear
signal of NP.

\begin{figure}
\center{\includegraphics[width=.7\textwidth]{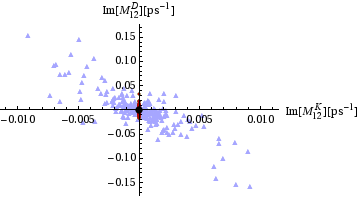}}
\caption{\it $\im(M_{12}^D)$ as a function of $\im(M_{12}^K)$. The red (dark
  grey) points fulfil all existing $K$ and $B$ physics constraints. ({The 
  two} solutions for $((M_{12}^D)_\text{SM},(\Gamma_{12}^D)_\text{SM})$ cannot
  be distinguished in this plot, the green points are covered by the red points.)
  For the light blue (grey) triangular points, the constraint from $\eps_K$ has
  been omitted.\label{fig:ImMK12-ImMD12}}
\end{figure}
Next in Fig.\ \ref{fig:ImMK12-ImMD12} we show the correlation between $\im (M_{12}^K)$ and $\im(M_{12}^D)$. We find a certain correlation between these two quantities, but as expected from the discussion in Section \ref{sec:xi}, this correlation is not a strict one due to the $\ord(\lambda)$ corrections to $\xi_i^{(D)}$ relative to $\xi_i^{(K)}$. Consistent with our previous results, we see also from this figure that including the experimental constraint from $\eps_K$ excludes very large values for $\im (M_{12}^D)$.

\newsection{\boldmath \cp~Asymmetries in $D$ Decays}
\label{sec:CPV}

From the first time they entered the stage of fundamental physics through the discovery of 
$K_L \to \pi^+\pi^-$, \cp~studies have demonstrated their power to reveal subtle dynamical 
features. We have good reason to expect that they will again reveal the intervention of New Physics. One should keep two facts in mind: (i) Baryogenesis requires dynamics beyond the SM  \cp~violation. (ii) With the SM providing one amplitude, a \cp~asymmetry can be linear in a 
New Physics amplitude thus exhibiting a relatively high sensitivity to the latter.

\subsection{General Formalism}

The time evolution of initially pure $D^0$ and $\bar D^0$ states, respectively, can be obtained from solving \eqref{eq:DGL} and is given by
\bea\label{eq:D0phys}
|D^0(t)\rangle &=& f_+(t)|D^0\rangle - \frac{q}{p} f_-(t) |\bar D^0\rangle\,,\\\label{eq:D0barphys}
|\bar D^0(t)\rangle &=& - \frac{p}{q} f_-(t)|D^0\rangle +f_+(t) |\bar D^0\rangle\,,
\eea
where
\bea
f_+(t) &=& e^{-i \bar M t} e^{-\bar\Gamma t/2} \cos Qt\,,\\
f_-(t) &=& i e^{-i \bar M t} e^{-\bar\Gamma t/2} \sin Qt\,,
\eea
with $q/p$ given in \eqref{qoverp}, $\bar M = (M_1+M_2)/2$ and
\beq
Q=  \sqrt{(M_{12}^D-\frac{i}{2}\Gamma_{12}^D)({M_{12}^D}^*-\frac{i}{2}{\Gamma_{12}^D)}^*} = \frac{1}{2}(\Delta M_D-\frac{i}{2}\Delta\Gamma_D)\,.
\eeq

From \eqref{eq:D0phys}, \eqref{eq:D0barphys} we find for the time-dependent decay rates of $D^0(t)$, $\bar D^0(t)$ to a final state $f$:
\bea\label{eq:Df}
\Gamma(D^0(t) \to f) &=& \left|T(D^0\to f)\right|^2 e^{-\bar\Gamma t} \bigg[
\frac{1}{2}\left(1+|\lambda_f|^2\right) \cosh\frac{\Delta\Gamma_Dt}{2}\\
&&{}+ 
\frac{1}{2}\left(1-|\lambda_f|^2\right) \cos\Delta M_Dt
-\sinh\frac{\Delta\Gamma_Dt}{2}\re\lambda_f+\sin\Delta M_Dt\im\lambda_f
\bigg]\,,\nn\\\label{eq:Dbarf}
\Gamma(\bar D^0(t) \to f) &=& \left|(T(\bar D^0\to f)\right|^2 e^{-\bar\Gamma t} \Bigg[
\frac{1}{2}\left(1+\left|\frac{1}{\lambda_f}\right|^2\right) \cosh\frac{\Delta\Gamma_Dt}{2}\\
&&{}+ 
\frac{1}{2}\left(1- \left|\frac{1}{\lambda_f}\right|^2\right) \cos\Delta M_Dt
-\sinh\frac{\Delta\Gamma_Dt}{2}\re\frac{1}{\lambda_f}+\sin\Delta M_Dt\im\frac{1}{\lambda_f}
\Bigg]\,,\nn
\eea
where {we dropped the overall phase space factors and} defined
\beq\label{eq:lf}
\lambda_f = \frac{q}{p}\frac{T(\bar D^0\to f)}{T( D^0 \to f)}\,.
\eeq
{These general formulae agree with those of Dunietz and Rosner 
\cite{Dunietz:1986vi} after their
 definitions of $\Delta M$ and $\Delta\Gamma$ are adjusted to ours in 
\eqref{eq:DM} and \eqref{eq:DG}.}

From these results, one can easily obtain the \cp~asymmetries
\beq
\frac{\Gamma(D^0(t) \to f) -\Gamma(\bar D^0(t) \to  f)}{\Gamma(D^0(t) \to f) + \Gamma(\bar D^0(t) \to f)}
\eeq
where $f$ is a CP eigenstate 
\beq
\cp |f\rangle = \eta_f |f\rangle \,,\qquad \eta_f=\pm 1\,.
\eeq
We find
\beq\label{eq:asymgen}
\frac{\Gamma(D^0(t) \to f) -\Gamma(\bar D^0(t) \to  f)}{\Gamma(D^0(t) \to f) + \Gamma(\bar D^0(t) \to  f)} = \frac{F(-)}{F(+)+\cosh\frac{\Delta\Gamma_Dt}{2}+\cos\Delta M_Dt }\,,
\eeq
where we have introduced the function
\bea
F(\pm) &=& \frac{1}{2}\left(\left|\frac{q}{p}\right|^2 \pm \left|\frac{p}{q}\right|^2 \right) \left( \cosh\frac{\Delta\Gamma_Dt}{2}-\cos\Delta M_Dt\right) \nn\\
&& {} - \left[\left(\left|\lambda_f\right|\pm \left|\frac{1}{\lambda_f}\right|\right)\cos2\varphi_f \sinh\frac{\Delta\Gamma_Dt}{2}
- \left(\left|\lambda_f\right|\mp \left|\frac{1}{\lambda_f}\right|\right)\sin2\varphi_f\sin\Delta M_Dt \right],\qquad
\eea
and
\beq
\varphi_f = \frac{1}{2}\arg\left(\lambda_f\right)\,.
\eeq
{We emphasize that the phase $\varphi_f$ is phase convention independent
as it depends only on the relative phase between $q/p$ and 
$T(\bar D^0\to f)/T(D^0\to f)$ in \eqref{eq:lf}.}
{The expression  \eqref{eq:asymgen} with $F(\pm)$ given above generalizes 
the {well-known formula from the $B$ system} to include the effects of $\Delta\Gamma$ and
$|q/p|\not=1$.}

For practical purposes, as $x_D,y_D\ll 1$ it is sufficient to consider the \cp~asymmetry in the limit of small $t$. Then \eqref{eq:asymgen} reduces to 
\bea\label{eq:smallt}
&& \frac{\Gamma(D^0(t) \to f) -\Gamma(\bar D^0(t) \to  f)}{\Gamma(D^0(t) \to f) + \Gamma(\bar D^0(t) \to  f)} \nn\\
&&\hspace{1cm} = -  \left[y_D \left(\left|\lambda_f\right|- 
\left|\frac{1}{\lambda_f}\right|\right)\cos2\varphi_f 
- x_D \left(\left|\lambda_f\right|+ \left|\frac{1}{\lambda_f}\right|\right)\sin2\varphi_f \right]\frac{t}{2\bar\tau_D}\,,
\eea
where $\bar\tau_D=1/\bar\Gamma$.

{In the {case of a non-negligible 
CP phase $\xi_f$ in the decay amplitude $T(D^0\to f)$, but  $|T(D^0\to f)|= |T(\bar D^0\to f)|$,} $\lambda_f$ simplifies
to}
\beq
\lambda_f = \eta_f \frac{q}{p} e^{-i2\xi_f},\qquad 
\left|\lambda_f\right|=\left|\frac{q}{p}\right|.
\eeq
{Moreover, if in the adopted phase convention (like CKM convention) the 
phase $\xi_f$ is negligible as assumed in what follows, we have}
\beq
\lambda_f = \eta_f \frac{q}{p} =
\eta_f\left|\frac{q}{p}\right|e^{i2\tilde\varphi}\,, \qquad {\tilde\varphi = \frac{1}{2}\arg\frac{q}{p}}\,.
\eeq
{We then find} 
\be
\frac{\Gamma(D^0(t) \to f) - \Gamma(\bar D^0(t) \to f)}
{\Gamma (D^0(t) \to f) + \Gamma(\bar D^0(t) \to f)} 
\equiv S_{f} \frac{t}{ 2\overline\tau _D} \,,
\label{eq:GASYM}
\ee
{where we defined in analogy with the $B$ system}
\be
S_f\simeq - \eta_f  \left[ y_D\left(\left| \frac{q}{p}\right| -\left| \frac{p}{q}\right|   \right)\cos2\tilde\varphi -
x_D \left(\left| \frac{q}{p}\right| +\left| \frac{p}{q}\right|   \right) \sin2\tilde\varphi \right] \,.
\label{eq:Sf}
\ee
{Note that in the $B$ system $y \ll x$ and $|q/p|\simeq1$, so that the above result simplifies considerably in the case of the CP asymmetries $S_{\psi K_S}$ and $S_{\psi\phi}$ in the $B_d$ and $B_s$ systems, respectively.}

{Finally we introduce the semileptonic asymmetry}
\beq\label{eq:ASL}
a_\text{SL}(D^0) \equiv  \frac{\Gamma (D^0(t) \to \ell^-\bar\nu K^{+(*)}) - \Gamma (\bar D^0 \to \ell^+\nu K^{-(*)})}
{\Gamma (D^0(t) \to \ell^-\bar\nu K^{+(*)}) + \Gamma (\bar D^0 \to \ell^+\nu K^{-(*)})}= 
\frac{|q|^4 - |p|^4}{|q|^4 + |p|^4}
\approx 2\left(\left|\frac{q}{p}\right|-1\right)
\eeq
{which represents \cp~violation in ${\cal L}(\Delta C=2)$. In writing the last
expression, we assumed that $|{q}/{p}|-1$ is much smaller
than unity.}

\subsection{Correlations}
\label{sec:corr}

Having all these formulae at hand we can derive two {interesting} correlations.
Following the presentation in \cite{Grossman:2009mn}, we find 
\be\label{eq:varphi}
\sin^22\tilde\varphi=\frac{x_D^2(1-|q/p|^2)^2}{x_D^2(1-|q/p|^2)^2+y_D^2(1+|q/p|^2)^2},
\ee
where in the phase conventions adopted in \eqref{eq:CPC} $\tilde\varphi = 1/2 \arg(q/p)$.
In the limit $\big| |q/p|-1 \big| \ll 1$, $x_D\sim y_D$ \eqref{eq:varphi} reduces to\footnote{We note though that the present data \eqref{eq:qpexp} allow still for a sizable deviation of $|q/p|$ from unity.} \cite{Grossman:2009mn}
\be
\sin 2\tilde\varphi=\frac{x_D}{y_D}\left(1-\left|\frac{q}{p}\right|\right)\,,
\ee
{where the sign ambiguity in taking the square root of \eqref{eq:varphi} can be resolved numerically.}
Using then \eqref{eq:Sf} and \eqref{eq:ASL} we find for $\xi_f=0$
\be
\label{eq:CORR1}
S_f=-\eta_f \frac{x_D^2+y_D^2}{y_D} a_\text{SL}(D^0)\,,
\ee
A similar correlation is familiar from the $B_s$ system \cite{Ligeti:2006pm,Blanke:2006ig,Grossman:2009mn} and we recall it in
Appendix~\ref{app:Bs}  using the formulation presented above.

{The violation of the relation \eqref{eq:CORR1} in future
experiments would imply the presence of direct CP violation
at work  \cite{Grossman:2009mn}. In the presence of a significant phase $\xi_f$ we
find
\be\label{eq:CORR2}
S_f=-\eta_f \left[ \cos2\xi_f \frac{x_D^2+y_D^2}{y_D}a_\text{SL}(D^0)
+2x_D \sin2\xi_f\right]\,.
\ee}

A similar comment applies to the correlation in $B_s$ physics that we discuss in
 Appendix~\ref{app:Bs}.

\section{LHT Results}\label{sec:CPV-LHT}

\subsection{SM Expectations}

It is generally understood that CKM dynamics can generate {\em direct} \cp~violation in Cabibbo suppressed 
modes. For --- in the Wolfenstein parameterisation of the CKM matrix ---  $V_{cs}$ contains a weak 
phase of order $\lambda^4$ and on the Cabibbo suppressed level there can be two different, yet coherent amplitudes. The same weak phase can also induce 
\cp~violation in ${\cal L}(\Delta C=2)$ {\cite{Bigi:2000wn,Bianco:2003vb}}. Yet because these effects are largely shaped by long distance dynamics, we cannot go beyond saying  that while SM dynamics generate \cp~violation in {some} charm transitions, it should happen below the $0.1$ \% level. Since it seems 
unlikely that the experimental uncertainties can be suppressed below that regime in the near future, we will ignore in our subsequent discussion  SM \cp~violating effects.

\subsection{LHT Scenarios}

Our findings above strongly suggest that LHT dynamics presumably generate \cp\ asymmetries in many 
different channels and Cabibbo levels, both of the indirect and direct variety. Exploring this rich experimental landscape in a comprehensive way will be left to a future paper. Here we will focus on the simplest case, namely on {\em indirect} \cp~violation entering through ${\cal L}(\Delta C = 2)$. Its impact 
on decay rates can be expressed through two types of observables: 
\beq 
\left| \frac{q}{p}\right| - 1 \; , 
\eeq
which describes \cp~violation in $D^0 - \bar D^0$ oscillations and 
\beq 
\im\lambda_f\,, 
\eeq
reflecting the interplay between \cp~violation in the oscillations and the transition to a final state $f$. 
These two types of observables can be probed by analysing the time evolution of   
$D^0 \to K_SK^+K^-$, $K^+K^-$, $\pi^+\pi^-$, $K_S\pi^+\pi^-$, $\ell \nu K^{(*)}$. As stated above, 
in this paper we will assume that New Physics does not affect the direct decay amplitudes for those transitions in any appreciable way.

From the definition of $q/p$ one easily obtains {(see also \cite{Bigi:2009jj})}
\beq 
\left| \frac{q}{p}\right| ^4 = \frac{1 +\left| \frac{\Gamma_{12}^D}{2M_{12}^D}\right|^2 + 
\left| \frac{\Gamma_{12}^D}{M_{12}^D}\right|\sin 2\varphi_{12}}
{1 +\left| \frac{\Gamma_{12}^D}{2M_{12}^D}\right|^2  -
\left| \frac{\Gamma_{12}^D}{M_{12}^D}\right|\sin2\varphi_{12}}\,,
\eeq
where $\varphi_{12}$ has been defined in \eqref{eq:phi12}.
With $M_{12}^D  = (M_{12}^D)_\text{SM} + (M_{12}^D)_\text{LHT}$ and $\Gamma_{12}^D = (\Gamma_{12}^D)_\text{SM}$ as determined above we can evaluate $|q/p|$. Likewise for $\im\lambda_f$, provided the phase $2\xi_f$ from the ratio of decay amplitudes can be neglected. Below we go through a typical list of transitions, {where} these \cp~observables can be probed.

\subsubsection[$D^0 \to \ell \nu K^{(*)}$]{\boldmath $D^0 \to \ell \nu K^{(*)}$}
\label{SLCPV}

Because of the SM selection rule, `wrong'-sign leptons --- 
$D^0 \to\ell^- \bar \nu K^{+(*)}$, $\bar D^0 \to\ell^+  \nu K^{-(*)}$ --- are theoretically the cleanest signature for oscillations. Having  such wrong sign leptons one can search for a difference in them, expressed through $a_\text{SL}(D^0)$ in \eqref{eq:ASL},
which represents \cp~violation in ${\cal L}(\Delta C=2)$. While the rate of wrong sign lepton production oscillates with time, this \cp~asymmetry does not.

\begin{figure}
\center{\includegraphics[width=.7\textwidth]{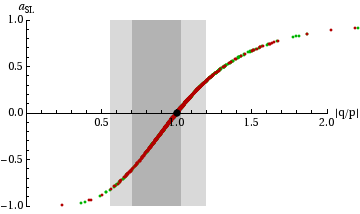}}
\caption{\it $a_\text{SL}(D^0)$ as a function of $|q/p|$. The red (darker grey) and green (lighter grey) points correspond to the two solutions for $((M_{12}^D)_\text{SM},(\Gamma_{12}^D)_\text{SM})$. The dark and light grey bands correspond to the experimental $1\sigma$ and $2\sigma$ ranges for $|q/p|$, as given in \eqref{eq:qpexp}.\label{fig:qp-aSL}}
\end{figure}

{In Fig.\ \ref{fig:qp-aSL} we show $a_\text{SL}(D^0)$ as a function 
of $|q/p|$. We observe that almost any value for $a_\text{SL}(D^0)$
can be generated by the LHT model, but the existing measurements for
$|q/p|$ constrain the possible range for $a_\text{SL}(D^0)$.
}
The important point to note here is the following: We know already from the data that the production 
probability of wrong-sign leptons is very low as expressed by $\frac{x_D^2 + y_D^2}{2}$; yet this 
still allows for a sizable or even large \cp~asymmetry there: in the LHT model one
could get numbers as large as 
\beq \label{eq:aSL-qpexp}
-0.8 \lsim a_\text{SL}(D^0) \lsim + 0.3
\eeq
{restricted only {by the measured bounds on} $|q/p|$ (the above range
corresponds to the current experimental $2\sigma$ range in \eqref{eq:qpexp}).}

This finding is relevant even when one cannot measure $a_\text{SL}(D^0)$ directly as is the case in hadronic collisions. For \cp~asymmetries in nonleptonic $D^0$ transitions depend, as we will discuss below, on the same quantity $|q/p|$ which underlies $a_\text{SL}(D^0)$, so that stringent correlations between the various asymmetries exist {(see Sect.\ \ref{sec:CPV})}.

\subsubsection[$D^0 \to K_S \phi$, $K_S K^+K^-$, $K_S\pi^+\pi^-$]{\boldmath $D^0 \to K_S \phi$, $K_S K^+K^-$, $K_S\pi^+\pi^-$}
\label{NLCPV}

As already indicated above we feel quite safe in ignoring {\em direct} \cp~violation for 
Cabibbo favoured modes. The theoretically simplest channels would be 
$D^0 \to K_S\pi^0$, $K_S\eta$, $K_S \eta^{\prime}$ --- alas experimentally they 
are anything but simple. In a hadronic environment they seem to be close to impossible. The next best 
mode is
\beq 
D^0 \to K_S\phi \to K_S[K^+K^-]_{\phi} \; , 
\eeq
which (apart from a doubly Cabibbo suppressed transition $D^0 \to K^0\phi$) is given by a single 
isospin amplitude. The {\em strong} phase thus drops out from the ratio 
$\frac{T(\bar D^0 \to K_S\phi)}{T(D^0 \to K_S\phi)}$, while their SM weak 
phase can be ignored at first.\footnote{The KM weak {phases} in ${T(\bar D^0 \to K_S\phi)}/{T(D^0 \to K_S\phi)}$ and ${q}/{p}$ actually cancel to good accuracy.} 
{Therefore we can use \eqref{eq:Sf} with $\eta_{K_S\phi}=-1$,}
in {\em qualitative}  analogy to $B_d \to \psi K_S$. The effect will be much smaller of course, 
since the oscillations proceed much more slowly and a priori one cannot ignore the impact of 
$y_D \neq 0$ and 
$|q/p| \neq 1$. Furthermore the experimental signature of $\phi$ is not nearly as 
striking as that of $\psi$. One has to extract it from the $K_SK^+K^-$ final state and distinguish it 
from final states like $K_Sf^0$. The latter is particularly important, since the \cp~parities of 
$K_Sf^0$ and $K_S\phi$ are opposite. Therefore these final states would have to exhibit \cp~asymmetries of equal size, yet opposite sign. Ultimately one has to and can perform a 
\cp~analysis of the full Dalitz plot for $K_SK^+K^-$; describing it is beyond the scope of this paper.

\begin{figure}
\center{\includegraphics[width=.7\textwidth]{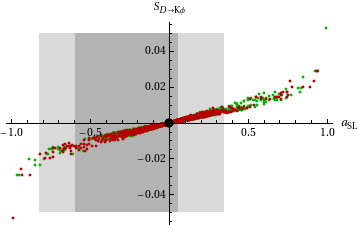}}
\caption{\it Correlation between the \cp~asymmetries $a_\text{SL}(D^0)$ and $S_{D\to K_S\phi}$. The red (darker grey) and green (lighter grey) points correspond to the two solutions for $((M_{12}^D)_\text{SM},(\Gamma_{12}^D)_\text{SM})$. {The dark and light grey bands correspond to the experimental $1\sigma$ and $2\sigma$ ranges for $|q/p|$, as given in \eqref{eq:qpexp}.} \label{aSL-SDKphi}}
\end{figure}

{In Fig.\ \ref{aSL-SDKphi} we show the correlation between $S_{D\to K_S\phi}$ and $a_\text{SL}(D^0)$. While we can see that a priori LHT dynamics could generate values for $S_{D\to K_S\phi}$ as large as $\pm 0.05$, the experimental constraint on $|q/p|$ in \eqref{eq:qpexp} and consequently on $a_\text{SL}(D^0)$ in \eqref{eq:aSL-qpexp}, {displayed by the grey band in the plot}, implies an allowed range
\be
-0.02 \lsim S_{D\to K_S\phi} \lsim + 0.01\,,
\ee
due to the strong correlation between the two CP asymmetries. We observe that for realistic values of $a_\text{SL}(D^0)$, as given in \eqref{eq:aSL-qpexp}, the strict correlation between these two CP asymmetries is linear to an excellent approximation, with the gradient given by $(x_D^2+y_D^2)/y_D\sim 0.02$, as derived analytically in \eqref{eq:CORR1}. As already discussed in Section \ref{sec:corr}, the violation of the correlation in question
would signal {the presence of direct CP violation in the $D^0\to K_S\phi$ decay.}}

Another suitable channel on the Cabibbo allowed level is $D^0 \to K_S\pi^+\pi^-$. One starts 
with resonant final states $D^0 \to K_S\rho^0$ etc. and then proceeds to a Dalitz plot analysis. There is 
an additional complication though: in general one has to deal with more than one isospin 
amplitude. {Therefore we leave a detailed study of this channel in the LHT model for future work.}

\subsubsection[$D^0 \to K^+\pi^-$]{\boldmath $D^0 \to K^+\pi^-$}

Neither CKM nor, it seems, LHT dynamics can generate direct \cp~violation for this doubly Cabibbo 
suppressed mode. Any \cp~asymmetry in this mode has to be of the indirect variety \cite{Bigi:ICHEP,Blaylock:1995ay} involving 
oscillations (unless there is still another source of \cp~violation \cite{D'Ambrosio:2001wg}). Its sensitivity to oscillation effects 
is actually enhanced, since the direct decay amplitude is considerably reduced by $\sim \lambda^2$. 
Accordingly it already figures prominently in the data base for $D^0$ oscillations. 

\subsubsection[$D^0 \to K^+K^-$, $\pi^+\pi^-$]{\boldmath $D^0 \to K^+K^-$, $\pi^+\pi^-$}

LHT dynamics can generate direct \cp~violation here due to Penguin diagrams. Evaluating their impact 
in a reliable way remains a task to be done. Both transitions can exhibit time dependent \cp~violation driven by the oscillation phase 2$\varphi_{12}$. 

\newsection{\boldmath Impact of LHT Dynamics on $K$ and $B$ Decays}
\label{sec:KB}

While Little Higgs models follow only one among many routes towards \NP, LHT dynamics create non-trivial connections between what might emerge in high 
$p_t$ collisions at the LHC and flavour dynamics in principle --- SUSY models do that as well ---, 
but also in practice due to its relative paucity in additional model
parameters. Here we have discussed its impact on the transitions of neutral $D$
mesons. Yet it creates intriguing effects also in kaon and $B$ decays as
described in detail in \cite{Hubisz:2005bd,Blanke:2006sb,Blanke:2006eb,Blanke:2007db,Blanke:2007wr,Blanke:2008ac,LHT-update}

{In Fig.\ \ref{Spsiphi-qp} we plot the \cp~asymmetry in 
$B_s \to \psi \phi$ ($S_{\psi\phi}$) against $|q/p|$ in the $D^0-\bar D^0$ system. 
In the LHT model, $S_{\psi\phi}$ can easily reach values
between $-0.2$ and $+0.3$, {i.\,e.
considerably larger than its SM prediction that is Cabibbo suppressed \cite{Bigi:1981qs}}; even larger values (up to $+0.6$) are possible for
some points in parameter space \cite{Blanke:2006sb,Blanke:2008ac,LHT-update}. We observe a cross-like structure in the plot, meaning that while either $S_{\psi\phi}$ or $|q/p|$ in the $D^0-\bar D^0$ system can deviate significantly from their SM predictions 0.04 and 1, respectively, it is unlikely to observe large deviations
from the SM values in both quantities simultaneously. Therefore if the present hints for a large non-SM value of $S_{\psi\phi}$ \cite{Aaltonen:2007he,:2008fj,Brooijmans:2008nt} will be confirmed by more accurate data, LHT dynamics will probably {\it not} lead to large CP violating effects in $D^0-\bar D^0$ oscillations, albeit visible effects are still possible. On the other hand, if {eventually} $S_{\psi\phi}$ will turn out to be SM-like, the road towards spectacular LHT effects in $D^0-\bar D^0$ CP violation will {still} be open.

\begin{figure}
\center{\includegraphics[width=.7\textwidth]{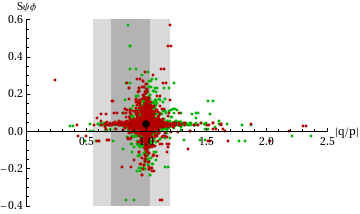}}
\caption{\it $S_{\psi\phi}$ as a function of $|q/p|$. The red (darker grey) and green (lighter grey) points correspond to the two solutions for $((M_{12}^D)_\text{SM},(\Gamma_{12}^D)_\text{SM})$.\label{Spsiphi-qp}}
\end{figure}

Fig.\ \ref{KL-qp} shows $Br(K_L \to \pi^0 \nu \bar \nu)$ plotted
against $|q/p|$ in the $D^0-\bar D^0$ system. In order to study the impact of the  value of $S_{\psi\phi}$ in this correlation, we show it, in the left panel of Fig.\ \ref{KL-qp},
 for SM-like values of $S_{\psi\phi}$ ($<0.05$) and, in the right panel of Fig.\ \ref{KL-qp},
for larger values of $S_{\psi\phi}$ ($\ge0.1$).  
$Br(K_L \to \pi^0 \nu \bar \nu)$ can reach values up to
$1.5\cdot 10^{-10}$ (more than four times the SM expectation) \cite{LHT-update}.
For larger values of $S_{\psi\phi}$ (right hand side
of the plot) very large enhancements of $Br(K_L \to \pi^0 \nu \bar \nu)$
are not observed\footnote{It should be
noted, however, that the right hand side of Fig.\ \ref{KL-qp} contains
considerably less {parameter} points than the left hand side.}, this shows that simultaneous large LHT effects
in CP violating $K$ and $B$ decays are unlikely \cite{Blanke:2006eb,Blanke:2008ac,LHT-update}\footnote{The situation is in fact analogous to the one observed in Fig.\ \ref{Spsiphi-qp} and shows that simultaneous large effects in $B$ and in $K$ or $D$ decays are generally unlikely in the LHT model}.
On the other hand, from the distribution of points on the left hand side
we can see that large effects in $Br(K_L \to \pi^0 \nu \bar \nu)$
and in $|q/p|$ in the $D^0-\bar D^0$ system do not exclude each other, on the contrary:
In contrast to the cross-like structure in \ref{Spsiphi-qp}, we now observe an hourglass-like distribution of points, i.\,e. for points with large $Br(K_L \to \pi^0 \nu \bar \nu)$ extreme effects in $|q/p|$ up to $\approx 0.5$ or $\approx 2$ are much more likely than in the case of SM-like $Br(K_L \to \pi^0 \nu \bar \nu)$.
This shows the correlation between CP violation in the $K$ system
and in the $D$ system which was already apparent in Figs.\
\ref{fig:ReMD12-ImMD12} and \ref{fig:ImMK12-ImMD12}.
}
\begin{figure}
\begin{minipage}{7.5cm}
\center{\includegraphics[width=\textwidth]{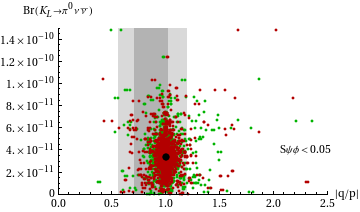}}
\end{minipage}\hfill
\begin{minipage}{7.5cm}
\center{\includegraphics[width=\textwidth]{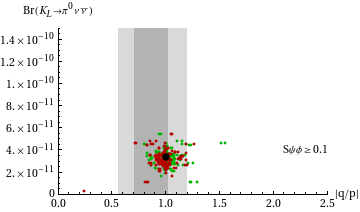}}
\end{minipage}
\caption{\it $Br(K_L\to\pi^0\nu\bar\nu)$ as a function of $|q/p|$, showing only points which predict $S_{\psi\phi}<0.05$ (left) and $S_{\psi\phi}\ge0.1$ (right). The red (darker grey) and green (lighter grey) points correspond to the two solutions for $((M_{12}^D)_\text{SM},(\Gamma_{12}^D)_\text{SM})$.\label{KL-qp}}
\end{figure}

\newsection{Summary and Outlook}
\label{sec:sum}

While the observed values of $\Delta M_D$ and $\Delta \Gamma_D$ might be generated by SM dynamics alone, $\Delta M_D$ could receive significant or even dominant contributions from New Physics. It would take a breakthrough in our control of nonperturbative effects to arrive at accurate {SM predictions for} $\Delta M_D$ and $\Delta \Gamma_D$. A more pragmatic approach to the 
interpretative conundrum posed by the observation of $D^0 - \bar D^0$ oscillations is to pursue a 
dedicated and comprehensive program of \cp~studies in $D^0$ transitions in particular. Oscillations 
can obviously reveal \cp~violation residing in ${\cal L}(\Delta C=2)$; in addition they can provide access to {\em direct} \cp~violation that otherwise would remain unobservable, namely in doubly Cabibbo 
suppressed modes. 

The fact that baryogenesis requires the intervention of New Physics with \cp~violation represents a 
generic motivation for the aforementioned program of \cp~studies. Here we have provided a much more 
specific one, namely one based on LHT models. Let us repeat: the construction of these models was guided by considerations based on electroweak rather than flavour dynamics --- yet they can have 
non-trivial consequences for the latter. Due to the paucity of their new parameters --- at least relative to their `competitors' --- they can create connections between the parameters describing the on-shell 
behaviour of their new quanta and \cp~asymmetries in $K$, $B$ and $D$ decays that might be of practical use.  What we have shown here is that LHT dynamics can generate sizable \cp~asymmetries 
in $D^0$ decays as expressed through $\im\lambda_f \neq 0$ and 
$\left| q/p \right| \neq 1$. The latter implies {among other things} that while $D^0 - \bar D^0$ oscillations produce only very few `wrong-sign' leptons, those might exhibit a sizable \cp~asymmetry. {More generally both of these
portals for CP violation can be studied in nonleptonic channels like $D^0 \to
K_SK^+K^-$, $K^+K^-$,
$\pi^+\pi^-$ and $K^+\pi^-$.}

{In summary the main analytic results of our paper can be found in the equations \eqref{eq:xi}, \eqref{eq:asymgen}, \eqref{eq:smallt}, \eqref{eq:Sf}, \eqref{eq:varphi}, \eqref{eq:CORR1}, \eqref{eq:CORR2} and \eqref{eq:CORRBs}. The corresponding phenomenological implications are discussed in Sections \ref{sec:CPV-LHT} and \ref{sec:KB}. More specifically in the present paper:
\bi
\item
We have presented a general formula \eqref{eq:asymgen} for the mixing induced time dependent CP asymmetry for decays into a CP eigenstate. Compared with the analogous formulae known from the $B$ system, \eqref{eq:asymgen} includes the effects of $\Delta\Gamma\ne0$ and $|q/p|\ne1$.
\item
Assuming the absence of {direct CP violation}  in the $\Delta C=1$ decay amplitudes, we have presented an expression for the CP asymmetry $S_f$ \eqref{eq:Sf} that generalises the familiar expressions for $S_{\psi K_S}$ and $S_{\psi\phi}$ to include the effects of $\Delta\Gamma\ne0$ and $|q/p|\ne1$.
\item
We have derived a correlation between $S_f$ and $a_{SL}(D^0)$  \eqref{eq:CORR1} that depends only on $x_D, y_D$ and $\eta_f$. A similar dependence has recently been pointed out in the case of $B_s-\bar B_s$ mixing in  \cite{Grossman:2009mn}. We confirm the latter {result and give it in appendix \ref{app:Bs}.}
\item
We have discussed the correlation between $D$ and $K$ decays in the spirit of the recent analysis in \cite{Blum:2009sk}, demonstrating that the LHT model exhibits {this correlation} very transparently. To this end the expression \eqref{eq:xi} turned out to be very useful.
\item
Analysing in detail the LHT model we have found observable CP violating effects in $D^0-\bar D^0$ oscillations well beyond anything possible with CKM dynamics. The correlation between $S_f$ and $a_\text{SL}(D^0)$, illustrated here for $f=K_S\phi$ will serve as a useful test (see Fig.\ \ref{aSL-SDKphi}) of the LHT dynamics. 
\item
We have identified a clear pattern of flavour violation predicted by the LHT model:
\bi
\item While either the CP asymmetry $S_{\psi\phi}$ in $B_s-\bar B_s$ mixing or $|q/p|$ in the $D^0-\bar D^0$ system can deviate significantly from their SM predictions, it is unlikely to observe large deviations from the SM values in both quantities simultaneously. The improved measurements of $S_{\psi\phi}$ at the Tevatron and the LHC in the coming years will therefore have a large impact on the possible size of CP violating effects in $D$ decays within the LHT model.
\item
The strong correlation between the $K$ and $D$ systems implies that large \NP~effects in $K$ and $D$ decays are possible simultaneously.
\item
On the other hand simultaneous large effects in $K$ and $B$ decays are unlikely \cite{Blanke:2006eb,Blanke:2008ac,LHT-update}. The latter property has also been pointed out recently in the context of RS models with custodial protection \cite{Blanke:2008zb,Blanke:2008yr}.
\ei
\ei

Finally the analysis presented
here can also be viewed as a proof of principle in two ways:
\begin{enumerate}
\item Charm decays might reveal the intervention of dynamics that so far has
remained hidden.
\item While the CP phenomenology of $K \to \pi \nu \bar \nu$, $B_s \to \psi
\phi$ and of
$D^0$ decays has some overlap, it is also fully complementary and its dedicated
study thus mandatory: new dynamics that might hardly affect $B_s \to \psi \phi$
can leave an identifiable footprint in $K \to \pi \nu \bar \nu$ and $D^0$
decays.
\end{enumerate}
}

  
\subsubsection*{Acknowledgements} 
I.B. thanks the Excellence Cluster of
the TUM for the wonderful hospitality afforded to him last year, when this work
was begun.
A.J.B. would like to thank the theory group of the Physics Department
of Cornell University for great hospitality during the final stages
of this paper. He would also like to thank Robert Fleischer for useful discussions.
This research was supported by the NSF under the grant number PHY-0807959, 
the Deutsche
Forschungsgemeinschaft (DFG) under contract BU 706/2-1, the DFG Cluster of
Excellence `Origin and Structure of the Universe' and by the German
Bundesministerium f{\"u}r Bildung und Forschung under contract 05HT6WOA.


\begin{appendix}

\newsection{\boldmath CP Violation in $B_s-\bar B_s$ Mixing}\label{app:Bs}

The general formulae for CP asymmetries discussed in Sect.\ \ref{sec:CPV} can be applied to the $B_s$ meson system as well. We recall that in the latter system $|q/p|=1$ with good accuracy, and in addition $y\ll x$. {Using \eqref{eq:varphi}  we find
\be\label{eq:CORRBs}
a^s_\text{SL} = -2 \left|\frac{y_s}{x_s}\right| \frac{S_{\psi\phi}}{\sqrt{1-S_{\psi\phi}^2}}\,,
\ee
which agrees with the findings in version 3 of \cite{Grossman:2009mn}} {and represents an alternative derivation of the correlation found in \cite{Ligeti:2006pm,Blanke:2006ig}}. Note that
we used the {definition}
\be
a^s_\text{SL} = \frac{\Gamma (\bar B_s(t) \to \ell^+ X) - \Gamma (B_s(t) \to \ell^- X)}
{\Gamma (\bar B_s(t) \to \ell^+ X) + \Gamma ( B_s(t) \to \ell^- X)}\,,
\ee
and $S_{\psi\phi}$ is the coefficient of $\sin\Delta M_st$ in
\be
\frac{\Gamma (\bar B_s(t) \to \psi\phi) - \Gamma (B_s(t) \to  \psi\phi)}
{\Gamma (\bar B_s(t) \to  \psi\phi) + \Gamma ( B_s(t) \to  \psi\phi)}\,.
\ee
Further
\be
x_s = \frac{m_H-m_L}{\bar\Gamma_s}\,,\qquad y_s = \frac{\Gamma_H-\Gamma_L}{2\bar\Gamma_s}\,,
\ee
where we stress that our definition of $y_s$ differs by sign from the HFAG one. {Finally in determining the overall sign of \eqref{eq:CORRBs} we assumed $(y_s)_\text{SM} < 0$.}

\newsection{Numerical input}\label{app:input}
We use the following values of the experimental and theoretical
    quantities as input parameters:

\begin{center}
\begin{tabular}{|l|l|}
\hline
$\lambda=|V_{us}|= 0.226(2)$ & $G_F= 1.16637\cdot 10^{-5}\gev^{-2}$ \qquad {} \\
$|V_{ub}| = 3.8(4)\cdot 10^{-3}$ &  $M_W = 80.398(25) \gev$ \\
$|V_{cb}|= 4.1(1)\cdot 10^{-2}$
&  $\sin^2\theta_W = 0.23122$\\ 
$ \gamma = 78(12)^\circ $  &$m_{K^0}= 497.614\mev$\\\cline{1-1}
$\Delta M_K= 0.5292(9)\cdot 10^{-2} \,\text{ps}^{-1}$ & $m_{B_d}= 5279.5\mev$   \\
$|\eps_K|= 2.229(12)\cdot 10^{-3}$ & $m_{B_s} = 5366.4\mev$ \\\cline{1-1}
$\Delta M_d = 0.507(5) \,\text{ps}^{-1}$ &$m_{D^0}=1864.6\mev$\\\cline{2-2}
$\Delta M_s = 17.77(12) \,\text{ps}^{-1}$  & $\eta_1= 1.43(23)$ \\
$S_{\psi K_S}= 0.675(26)$ &  $\eta_3=0.47(4)$ \\\hline
$\bar m_c = 1.27(2)\gev$ & $\eta_2=0.577(7)$ \\
$\bar m_t = 162.7(13)\gev$ & $\eta_B=0.55(1)$ \\\hline
$F_K = 156(1)\mev$ & $F_{B_s} = 245(25)\mev$ \\
$\hat B_K= 0.75(7)$ & $F_{B_d} = 200(20)\mev$ \\\cline{1-1}
$\hat B_{B_s} = 1.22(12)$ & $F_{B_s} \hat B_{B_s}^{1/2} = 270(30)\mev$ \\
$\hat B_{B_d} = 1.22(12)$ & $F_{B_d} \hat B_{B_d}^{1/2} = 225(25)\mev$ \\\cline{1-1}
$F_D=212\mev$ & $\xi = 1.21(4)$ \\\cline{2-2}
$\hat B_{D} =1.17$ & $\kappa_L=2.31\cdot 10^{-10}$\\
\hline
\end{tabular}  
\end{center}

\end{appendix}

\providecommand{\href}[2]{#2}\begingroup\raggedright\endgroup


\end{document}